\documentclass[12pt]{article}

\usepackage{amssymb}
\usepackage{amsmath}
\usepackage{latexsym}
\usepackage{graphicx}

\catcode`@=11
\renewcommand{\section}{\@startsection{section}{1}{0pt}{\medskipamount}
{\medskipamount}{\large\bf}}
\numberwithin{equation}{section}
\catcode`@=12

\def\beq{\begin{eqnarray}}    
\def\eeq{\end{eqnarray}}      

\def\ln{\,\mbox{ln}\,}                  

\newcommand{\R}{{\mathbb R}}

\def\pa{\partial}                       

\def\im{\textrm{i}}

\def\diff{\textrm{d}}
\def\sfrac#1#2{{\textstyle\frac{#1}{#2}}}
\def\={\ =\ }

\begin{document}

\begin{titlepage}
\setcounter{page}{0}

\begin{center}

{\LARGE\bf Once again on the Gribov horizon}

\vspace{10mm}

{\Large Peter M. Lavrov$\,{}^{a,b}$$^{}\footnote{E-mail:
lavrov@tspu.edu.ru}$\quad and
\quad Olga V. Radchenko
$\,{}^{a}$$^{}\footnote{E-mail: radchenko@tspu.edu.ru}$ }

\vspace{8mm}

\noindent ${}^{a}${\em
Tomsk State Pedagogical University,\\
Kievskaya St.\ 60, 634061 Tomsk, Russia}

\vspace{2mm}

\noindent  ${{}^{b}}${\em
National Research Tomsk State  University,\\
Lenin Av.\ 36, 634050 Tomsk, Russia}

\vspace{20mm}

\begin{abstract}
\noindent
The gauge dependence problem existing in the original Gribov-Zwanziger theory
is discussed.
\end{abstract}

\end{center}

\vfill


\end{titlepage}


\section{Introduction}

\noindent The Gribov-Zwanziger (GZ) theory \cite{Zw1,Zw2} is an
attempt to formulate a non-perturbative approach to quantum theory
of non-Abelian gauge fields when one takes into account the Gribov copies
\cite{Gribov}. There exists very intense activity to study quantum
properties of the GZ theory itself as well as QCD dynamics within
this method (see review \cite{VZw} and references therein). Note
that the original formulation of the GZ theory is used the Landau
gauge only. Despite numerous studies of quantum properties,
the gauge dependence problem in
the framework of the GZ theory was not considered at all until
recently. The study of this problem in Yang-Mills theories assuming
the existence of the Gribov horizon functional beyond the Landau
gauge only was given in \cite{LLR} and then was extended for general
gauge theories in \cite{LRR}. It was shown that in general the
vacuum functional as well as  the effective action even on its extremals
depend on gauges. It indicates that physical observables may be
not introduced in a consistent way
as quantities not depending on gauges. In its turn it was found that there is a
strong restriction on the gauge dependence of Gribov horizon functional
when  the effective action on its extremals does not depend on gauges.
It gave hope to formulate the GZ theory in a consistent way as
a physical theory. Later studying the finite
field dependent BRST transformations in Yang-Mills theories
\cite{LL1} allowed to propose a form of the Gribov horizon
functional beyond the Landau gauge which satisfy the above mentioned
restriction \cite{LL2}.

In recent studies \cite{GMS,CFS} the Gribov
horizon functionals in the Coulomb gauge \cite{Zw3} and linear gauges were used
to construct the corresponding GZ theory. Unfortunately the authors
of these papers did not pay attention to the gauge dependence problem in the GZ
theory which was the main goal of  studies \cite{LLR} although
the form of the Gribov horizon functional in these gauges differs
from the functionals proposed
in \cite{LL2}. So we are forced to attract attention
to the problem  once again in this short notice restricting ourselves to the case
of the vacuum functional (partition function) which appears
in the GZ theory formulated beyond the Landau
gauge.
\\

\section{Gauge dependence problem}

\noindent
Partition function, $Z_{\psi_0}$, in the original GZ theory
is given by the following functional
integral
\beq
\label{ZGZ}
Z_{\psi_0}=\int\!{\cal
D}\phi\ \exp\Big\{\frac{\im}{\hbar}\big( S_0(A)+
s\psi_0(\phi)+M_{\psi_0}(A)\big)\Big\}
\eeq
where
\beq
S_0(A) \= -\frac{1}{4}\int\!\diff^d x\ F_{\mu\nu}^{a}F^{\mu\nu{}a}
\qquad\textrm{with}\qquad
F^a_{\mu\nu}\=\partial_{\mu}A^a_{\nu}-\partial_{\nu}A^a_{\mu}+
f^{abc}A^b_{\mu}A^c_{\nu}\  \label{clYM}
\eeq
is the action of Yang-Mills fields $A^a_{\mu}(x)$ with gauge group SU$(n)$ in $d$
space-time dimensions and $a=1,\ldots,n^2{-}1$,
$\mu=0,1,\ldots,d{-}1$, $f^{abc}$ are the (totally antisymmetric)
structure constants of the Lie algebra~$su(n)$, $\phi$ denotes the set of fields
$\bigl\{\phi^A\bigr\}\=\bigl\{A^a_{\mu}, B^a, C^a, {\bar C}^a\bigr\}$, $B^a$ are
the Nakanishi-Lautrup auxiliary fields as well as $C^a$ and ${\bar C}^a$ present the
Faddeev-Popov ghost and antighost fields, respectively, with the following distribution
of the Grassmann parities~$\varepsilon$ and ghost numbers~${\rm gh}$
$\varepsilon(C^a)=\varepsilon(\bar C)^a=1$,
$\varepsilon(A^a_\mu)=\varepsilon(B^a)=0$,
${\rm gh}(A^a_\mu)={\rm gh}(B^a)=0$, ${\rm gh}(C^a)=-{\rm gh}({\bar C}^a)=1$.
In (\ref{ZGZ}) $\psi_0(\phi)$ is gauge fixing functional
in the Landau gauge\footnote{Here and below we use the DeWitt's
condensed notations \cite{DeWitt}.}
\beq
\label{psi0}
\psi_0(\phi)={\bar C}^a\chi^a(A),\quad \chi^a(A)=\pa^{\mu}A^a_{\mu},
\eeq
$M(A)$ denotes the Gribov horizon functional \cite{Zw1}
\beq
\label{FuncM}
M_{\psi_0}(A)=\gamma^2\,f^{abc}A^b_{\mu}(K^{-1})^{ad}(A)f^{dec}A^{e\mu}
 + \gamma^4\,d(n^2{-}1)\ ,
\eeq
where $K^{-1}$ inverts the (matrix-valued) Faddeev-Popov operator $K^{ab}(A)$
\beq
\label{FPmat}
K^{ab}(A)\=\frac{\delta\chi^a(A)}{\delta A^c_{\mu}}D^{cb}_{\mu}\=
\partial^{\mu}D_{\mu}^{ab}\=
\delta^{ab}\partial^{\mu}\partial_{\mu}+ f^{acb}A^c_{\mu}\partial^{\mu}
\eeq
 and $\gamma\in\R$ is the so-called thermodynamic or Gribov
parameter~\cite{Zw1,Zw2}. Finally $s$ means the nilpotent BRST operator presenting the BRST
transformations
\beq
\label{BRSTtr}
\delta_{B} A_{\mu}^{a} = D^{ab}_{\mu}C^b\lambda\ ,\quad
\delta_{B} \bar{C}{}^a = B^a\lambda\ ,\quad
\delta_{B} B^a = 0\ ,\quad
\delta_{B} C^a = \sfrac12 f^{abc}C^bC^c\lambda
\label{BRSTGZred}
\eeq
in the form
\beq
\delta_B\phi^A=(s\phi^A)\lambda
\eeq
where $\lambda$ is a constant fermionic parameter. It should be noted that the Gribov horizon
functional (\ref{FuncM}) is not BRST invariant because
\beq
\label{sM}
sM_{\psi_0}(A)\=\gamma^2f^{abc}f^{cde}\bigl[2D^{bq}_{\mu}C^q(K^{-1})^{ad}-
f^{mpn}A^b_{\mu}(K^{-1})^{am}K^{pq}C^q(K^{-1})^{nd}\bigr]A^{e\mu}\ \neq\ 0\ .
\eeq
This point is very crucial for the GZ theory because the BRST symmetry and its nilpotency
play the fundamental role in a consistent formulation of the quantum theory of gauge fields
\cite{KO}.

The Gribov horizon
functional (\ref{FuncM}) is written in the non-local form although there exists a
local presentation based on an extended configuration space which is equipped  with
additional two doublets of fermionic and bosonic fields \cite{Zw2}. For our aims
it is enough to work with the original configuration space of fields $\phi$ and with the
non-local Gribov horizon functional. It seems quite natural to assume existence
of the Gribov horizon functional and its gauge dependence beyond the Landau gauge. Let
$\psi=\psi(\phi)$ be an admissible gauge which differs from the case of the Landau gauge
fixing functional $\psi_0=\psi_0(\phi)$ (\ref{psi0}) and $M_{\psi}(\phi)$
be a corresponding Gribov horizon functional. It should be noted that we don't know
an explicit expression for $M_{\psi}$ in arbitrary gauge $\psi$ but
it is not an obstacle to write the vacuum functional of the GZ theory in the
form of functional integral
\beq
\label{ZGZpsi}
Z_{\psi}=\int\!{\cal
D}\phi\ \exp\Big\{\frac{\im}{\hbar}\big( S_0(A)+
s\psi(\phi)+M_{\psi}(\phi)\big)\Big\}.
\eeq
It follows from (\ref{ZGZpsi}) that variation of $Z_{\psi}$ under infinitesimal
change of gauge fixing functional $\psi\rightarrow\psi+\delta\psi$
reads
\beq
\label{Zpsi}
\delta Z_{\psi}=\frac{i}{\hbar}\big(\langle s\delta\psi\rangle+
\langle\delta M_{\psi}\rangle\big)
\eeq
where the notation
\beq
\langle(...)\rangle=\int\!{\cal
D}\phi\ (...)\exp\Big\{\frac{\im}{\hbar}\big( S_0(A)+
s\psi(\phi)+M_{\psi}(\phi)\big)\Big\}
\eeq
is used and $M_{\psi+\delta\psi}=M_{\psi}+\delta M_{\psi}$.
If $\delta Z_{\psi}\neq 0$ or $\langle s\delta\psi\rangle+
\langle\delta M_{\psi}\rangle\neq 0$ then the vacuum functional
depends on gauge and the same is valid
for physical S-matrix due to the equivalence theorem \cite{KT}. In turn
a physical interpretation becomes impossible. There exists only one
possibility for the GZ theory to be a consistent one if
the following relation
\beq
\label{strong}
\langle s\delta\psi\rangle+
\langle\delta M_{\psi}\rangle=0
\eeq
holds. The restriction (\ref{strong}) seems very strong limitation on possible
structure of $M_{\psi}$ because of the fixed structure $s\delta\psi$ and
the known explicit form $M_{\psi}$ (\ref{FuncM}) in the Landau gauge (\ref{psi0}).
In general it is difficult to expect that the condition (\ref{strong}) may be satisfied.
Nevertheless a solution to the problem (\ref{strong}) has been proposed in \cite{LL2}.
\\

\section{Solution to the equation (\ref{strong})}

\noindent
Let us consider the change of variables
\beq
\label{trans}
\phi^A\;\rightarrow\; \phi^{'A}=\phi^A+(s\phi^A)\Lambda(\phi).
\eeq
in the functional integral (\ref{ZGZ}). Here $\Lambda(\phi)$
is an arbitrary odd Grassmann
functional of fields $\phi$. These transformations
are known as the field-dependent
BRST transformations in Yang-Mills theories \cite{JM,LL1} (see also resent
progress in this field \cite{BLT1,BLT2,BLT3,BBLT,BLT4,BLT5}).
The result of this change can be presented in the form \cite{LL2}
\beq
\label{ZGZtr}
Z_{\psi_0}\!=\!\int\!{\cal
D}\phi\ \exp\Big\{\frac{\im}{\hbar}\big[ S_0(A)+
s\psi_0(\phi)+i\hbar\ln(1+s\Lambda(\phi))+
M_{\psi_0}(A)+(sM_{\psi_0}(A))\Lambda(\phi)\big]\Big\}.
\eeq
Using arbitrariness in choice of functional $\Lambda(\phi)$ and the nilpotency of operator $s$
we can generate any admissible gauge $\psi(\phi)$ in the presentation (\ref{ZGZtr}).
Indeed following the paper \cite{LL1} let us consider the functional $\Lambda_{\psi}(\phi)$
\beq
\label{Lambda}
\Lambda_{\psi}(\phi)=(\psi{-}\psi_0)\bigl(s(\psi{-}\psi_0)\bigr)^{-1}
\left( \exp\left\{\frac{1}{\im\hbar}s(\psi{-}\psi_0)\right\}-1 \right).
\eeq
Substituting (\ref{Lambda}) ($\Lambda(\phi)=\Lambda_{\psi}(\phi)$) into (\ref{ZGZtr})
one obtains
\beq
\label{ZGZtr1}
Z_{\psi_0}\!=\!\int\!{\cal
D}\phi\ \exp\Big\{\frac{\im}{\hbar}\big[ S_0(A)+
s\psi(\phi)+ M_{\psi_0}(A)+(sM_{\psi_0}(A))\Lambda_{\psi}(\phi)\big]\Big\}.
\eeq
Note that $S_0(A)+s\psi(\phi)$ is the Faddeev-Popov action written in the gauge $\psi(\phi)$.
If we identify the functional $M_{\psi_0}(A)+(sM_{\psi_0}(A))\Lambda_{\psi}(\phi)$ with the Gribov
horizon functional $M_{\psi}(\phi)$ in the gauge $\psi(\phi)$,
\beq
\label{Mpsi}
M_{\psi}(\phi)=M_{\psi_0}(A)+(sM_{\psi_0}(A))\Lambda_{\psi}(\phi),
\eeq
then due to (\ref{Zpsi}) we arrive at the gauge independence of vacuum functional in the GZ theory
\beq
Z_{\psi_0}=Z_{\psi}.
\eeq
It should be stressed that the relation (\ref{Mpsi})
presents the definition of the Gribov horizon
functional beyond the Landau gauge leading to possibility of formulation of
the GZ theory in a consistent way. Derivation of this functional from
the first principles as it was made for the Gribov horizon functional in the Landau gauge
 \cite{Zw1,Zw2} may give an other answer.
 It looks that it is really happening in the case of
 the Coulomb gauge.
\\

\section{Coulomb gauge}

\noindent Consider the Coulomb gauge when the gauge fixing functions
$\chi^a(A)$ and the gauge fixing functional $\psi(\phi)$ are equal
to
\beq
\chi^a(A)=\pa^i A^a_i,\quad \psi(\phi)={\bar C}^a \pa^i
A^a_i.
\eeq
In this case
\beq
\label{spsi} \psi-\psi_0=-{\bar C}^a
\pa^0A^a_0,\quad s(\psi-\psi_0)=-B^a\pa^0 A^a_0+\pa^0{\bar
C}^aD^{ab}_0C^b.
\eeq
From (\ref{spsi}) it follows the
non-polynomial dependence  $M_{\psi}(\phi)$ (\ref{Mpsi}) on fields $C^a,
{\bar C}^a, B^a$. On the other hand in paper \cite{Zw3} the non-local form  of the
Gribov horizon functional in the Coulomb gauge has been written as
\beq
\label{MCG}
 M_{\psi}(A)=\gamma^2f^{\;abc}A^b_i(K^{-1})^{ad}f^{\;dec}A^e_i
-(d-1)(n^2-1)\gamma^4,
\eeq
where $(K^{-1})^{ab}$ is inverse to the Faddeev-Popov matrix in the Coulomb gauge
$K^{ab}=\pa^iD^{ab}_i$. The functional $M_{\psi}(A)$ (\ref{MCG}) differs essentially
from the functional $M_{\psi}(\phi)$ (\ref{Mpsi}) with relations (\ref{Lambda})
and (\ref{spsi}) because these functionals are defined in
different functional spaces. In our opinion it means that the original GZ theory
\cite{Zw1,Zw2,Zw3} meets
serious problem concerning a correct definition of physical observables.
\\

\section{Discussion}
\noindent In the present paper we have attracted attention to the
gauge dependence problem existing in the original GZ theory. For the
first time the problem has been discussed in paper \cite{LLR} where
analysis of gauge dependence was based on assumption that the Gribov
horizon functional exists beyond the Landau gauge. It was shown that
in general the partition function and the effective action on its extremals depend
on gauges. In our opinion it makes impossible physical interpretation of results
obtained within the original GZ theory. But in particular from the
analysis  given in \cite{LLR} it followed
existence of a special dependence of the Gribov horizon functional
on gauges when the effective action on its extremals will be gauge
invariant. Later such kind of the Gribov horizon functional has been
proposed \cite{LL2}. In the present paper we have constructed explicitly the form of
the Gribov horizon functional in the Coulomb gauge leading to the
gauge independence of the partition function. Unfortunately we have found that
this functional differs from the Gribov horizon functional in the
Coulomb gauge appearing in the GZ theory \cite{Zw3}.

Last but not least remark. Quite recently a reformulation of the original GZ theory
has been proposed \cite{SZw} to
improve  quantum properties of the theory concerning the BRST symmetry and its nilpotency.
In this connection it should be definitely noted that the reformulation
was performed with the help of change of variables which violates the
transformation laws of fields  under the Poincare group.
If we will consider the GZ theory \cite{Zw1,Zw2} formulated in accordance
with general principles of quantum field theory then
we should consider the new theory \cite{SZw}  as  ill-defined and vice versa.
These theories can not exist simultaneously as equivalent.
In particular, it means that results of quantum
calculations in the new theory have no any relations to the original GZ theory.
We conclude that the gauge dependence problem remains open in the original GZ theory.
\\

\section*{Acknowledgments}
\noindent
P. M. Lavrov thanks I.A.~Batalin and I.V.~Tyutin for useful discussions.
The work  is supported in part by the Presidential  grant 88.2014.2 for LRSS and by
the RFBR grant 15-02-03594.
\\


\end{document}